\documentclass[prd,amsmath,amssymb,reprint,showpacs,showkeys]{revtex4-1}
\pdfoutput=1 
\usepackage{graphicx}
\usepackage{dcolumn}
\usepackage{bm}

\begin{document}

\title[]{Impact of Recent RHIC Data on Helicity-Dependent  
Parton Distribution Functions}

\author{Emanuele R. Nocera}
\email{emanuele.nocera@physics.ox.ac.uk}
\affiliation{Rudolf Peierls Centre for Theoretical Physics, 
University of Oxford,
1 Keble Road, 
OX1 3NP, Oxford, United Kingdom}

\date{\today}

\begin{abstract}

I study the impact of recent measurements performed at the Relativistic Heavy 
Ion Collider on the determination of helicity-dependent parton 
distribution functions of the proton. Specifically, I consider: preliminary 
data on longitudinally single-spin asymmetries for $W$-boson production 
recorded by the STAR experiment during the 2013 run at a center-of-mass energy 
$\sqrt{s}=510$ GeV; data on longitudinally double-spin asymmetries for di-jet
production recorded by the STAR experiment during the 2009 run at  
$\sqrt{s}=200$ GeV; and data on longitudinally double-spin asymmetries for
neutral pion production recorded by the PHENIX experiment during a series of 
runs between 2006 and 2013 at $\sqrt{s}=200$ GeV and $\sqrt{s}=510$ GeV.
The impact of the new data is studied by applying Bayesian reweighting
to NNPDFpol1.1, the most recent global analysis of helicity-dependent parton 
distribution functions based on the NNDPF methodology. 
In comparsion to NNDPFpol1.1, a slightly 
more positive polarized gluon PDF, extended to somewhat smaller values of 
momentum fractions with narrower uncertainties, as well as a slightly more 
marked and more precise asymmetry between polarized up and down sea quarks 
are found in this study.

\end{abstract}



\maketitle

The Relativistic Heavy Ion Collider (RHIC) at Brookhaven National 
Laboratory~\cite{Harrison:2002es,*Harrison:2003sb} is the only
high-energy accelerator ever built which can collide polarized 
proton beams. This unique feature provides an unprecedented 
opportunity to investigate how the spin structure of the proton emerges from 
the interactions among its partonic constituents, in the established framework 
of perturbative Quantum Chromodynamics (QCD)~\cite{Bunce:2000uv}. 

One of the main goals of the RHIC spin physics 
program~\cite{Aschenauer:2015eha} is to investigate the amount of 
proton's polarization carried by gluons and sea quarks. 
This information is encoded in the helicity-dependent (or polarized) 
Parton Distribution Functions (PDFs)
\begin{equation}
\Delta f(x,\mu^2) \equiv f^{\uparrow}(x,\mu^2) - f^{\downarrow}(x,\mu^2)
\,\mbox{,}
\label{eq:polPDFs}
\end{equation} 
which are defined as the net densities of partons ($f$ denotes either
a quark, $q$, an antiquark, $\bar{q}$ or a gluon, $g$), with spin aligned 
along ($\uparrow$) or opposite ($\downarrow$) the polarization direction of 
the parent nucleon, carrying a fraction $x$ of its momentum. 
Following factorization~\cite{Collins:1989gx}, measured 
observables are built up as a convolution product between hard-scattering
matrix elements and PDFs, which encapsulate respectively the short- and 
long-distance parts of the interaction. The former can be computed 
in perturbative QCD, while the latter cannot. Nevertheless, perturbative QCD
corrections lead PDFs to depend on the factorization scale $\mu$, as
dictated by evolution equations~\cite{Gribov:1972ri,*Lipatov:1974qm,*Altarelli:1977zs,*Dokshitzer:1977sg}.
Provided a suitable set of measurements, usually from a variety of 
hard-scattering processes, and the corresponding kernels, one should then be 
able to determine the PDFs from the data in a global QCD analysis.

Some of the leading processes which are being investigated at RHIC include 
$W$-boson, jet and pion production in collisions where only one or both proton 
beams are longitudinally polarized. The corresponding measured observables are 
single- and double-spin asymmetries, 
\begin{equation}
A_{L}=\frac{\sigma^+-\sigma^-}{\sigma^++\sigma^-}
\ \ \ \ \ \ \ \ \ \ 
A_{LL}=\frac{\sigma^{++}-\sigma^{+-}}{\sigma^{++}+\sigma^{+-}}
\,\mbox{,}
\label{eq:asymmetry}
\end{equation}
defined as the ratio of the difference to the sum of cross sections with 
opposite polarizations of one proton beam ($+$ and $-$ denote the
proton beam polarization along or opposite its momentum in the center-of-mass
frame). 

On the one hand, the production of a jet or a pion 
arises mainly from gluon-initiated partonic 
subprocesses. Therefore, a measurement of the associated longitudinal spin 
asymmetry provides a powerful probe of  $\Delta g$, the polarized gluon PDF. 
Such a probe is particularly relevant, because $\Delta g$ is left almost 
unconstrained by measurements in fixed-target polarized lepton-nucleon 
scattering, where it enters the corresponding cross section only via 
higher-order corrections and scaling violations in the evolution.

On the other hand, the production of a $W$ boson is driven by a purely 
weak interaction which, because of its chiral nature, couples left-handed 
quarks with right-handed antiquarks. Therefore, a measurement of the associated
longitudinal spin asymmetry provides a clean probe of the flavor decomposition
into $\Delta u$, $\Delta\bar{u}$, $\Delta d$ and $\Delta\bar{d}$ PDFs. 
Such a probe is particularly relevant, because in fixed-target polarized
lepton-nucleon scattering quark and antiquark PDFs can be accessed only if
the process is semi-inclusive. One exploits correlations 
between the type of hadron produced in the final state and the flavor of its 
partonic progenitor. These correlations are expressed by Fragmentation 
Functions (FFs), which however are on the same footing as PDFs and can 
limit the accuracy in the description of the process.

In 2014, a first round of measurements recorded by 
RHIC~\cite{Aschenauer:2015eha} were combined with the data from fixed-target 
lepton-nucleon scattering in two separate next-to-leading order (NLO) global 
QCD analyses: DSSV14~\cite{deFlorian:2014yva} and 
NNPDFpol1.1~\cite{Nocera:2014gqa}. They upgraded the corresponding previous 
analyses, DSSV08~\cite{deFlorian:2008mr,*deFlorian:2009vb} and 
NNPDFpol1.0~\cite{Ball:2013lla}, which were based respectively on 
inclusive and semi-inclusive deep-inelastic scattering (DIS) data, supplemented
with a small amount of preliminary RHIC data up to 2005, and on inclusive DIS 
data only. In both the DSSV14 and NNPDFpol1.0 analyses, single-inclusive jet
production asymmetries, measured by the STAR experiment during the 
2005-2006~\cite{Adamczyk:2012qj} and 2009~\cite{Adamczyk:2014ozi} runs
at a center-of-mass energy $\sqrt{s}=200$ GeV, were included. Moreover,
neutral pion production asymmetries, measured by the PHENIX experiment during 
the 2006 run at $\sqrt{s}=62.4$ GeV~\cite{Adare:2008qb} and at
$\sqrt{s}=200$ GeV~\cite{Adare:2008aa}, as well as during the 2009 run at 
$\sqrt{s}=200$ GeV~\cite{Adare:2014hsq}, were included in the DSSV14 analysis;
$W$-boson production asymmetries, measured by the STAR experiment during 
the 2010-2011 runs at $\sqrt{s}=500$ GeV and $\sqrt{s}=510$ 
GeV~\cite{Adamczyk:2014xyw}, were included in the NNPDFpol1.1 analysis. 
Neutral pion production asymmetries were not included in NNPDFpol1.1 because,
like semi-inclusive fixed-target DIS data, they entail the knowledge of FFs.
Such a knowledge has significantly improved since then, at least for 
pions. Several new global QCD analyses of FFs have been performed
recently with a bunch of very precise data from 
$B$-factories~\cite{deFlorian:2014xna,*Hirai:2016loo,*Sato:2016wqj}, including
the first analysis based on the same methodology adopted in 
NNPDFpol1.1~\cite{Nocera:2017qgb}. It would then be interesting to revisit 
neutral pion production at RHIC in light of these new developments.

Both the DSSV14 and NNPDFpol1.1 analyses have shown the substantial impact
of RHIC measurements in improving our knowledge of polarized 
PDFs~\cite{Nocera:2015fxa}. Indeed, they 
revealed for the first time a sizable, positive polarization of gluons (in 
both analyses similar results were found despite the different data set 
included), and a sizable, positive asymmetry between polarized up and down sea 
quarks (in the NNPDFpol1.1 analysis, thanks to $W$-boson production data).

Since then, some new experimental measurements have become available from RHIC. 
In this contribution, I study the impact of some of them in the 
framework of a global analysis based on the NNPDF methodology.
Specifically, I focus on the following sets of data.

First, I consider preliminary measurements~\cite{Gunarathne:2017fnd} of the
single-spin asymmetry for $W$-boson production, $A_L^{W}$, recorded by the STAR 
experiment during the 2013 run at $\sqrt{s}=510$ GeV. The integrated
luminosity of the sample is $\mathcal{L}=246.2$ pb$^{-1}$. Asymmetries are
provided separately for $W^+$ and $W^-$ final states, which were reconstructed
from their $W^\pm\to e^\pm\nu$ decay channels, as a function of the decay 
electron/positron rapidity $\eta^{e^\pm}$. The data is presented in four
bins for $W^+$ and four bins for $W^-$ and is integrated over the 
electron/positron transverse momentum in the range $25$ GeV $<p_T^{e^\pm} < 50$
GeV. Using leading-order (LO) kinematics, one can expect this data to constrain
polarized quark and antiquark PDFs with $0.05\lesssim x \lesssim 0.4$ and 
$\mu\sim M_W$, where $M_W$ is the mass of the $W$ boson.
The uncertainties of the data are mainly statistical, though they are reduced 
by about $40\%$ with respect to the previous measurements taken during the 
2010-2011 runs. In the experimental analysis, they are combined with 
systematic uncertainties due to the unpolarized background dilutions.
Additional subdominant correlated systematics, originating from
the beam polarization ($3.3\%$ of the asymmetry) and from the relative 
luminosity (an offset of $0.004$ on the asymmetry), are provided separately. 

Second, I consider the first measurement~\cite{Adamczyk:2016okk} 
of the double-spin asymmetry for mid-rapidity di-jet production, 
$A_{LL}^{2{\rm jets}}$, recorded by the STAR experiment during the 2009 run at
$\sqrt{s}=200$ GeV. The integrated luminosity of the sample is 
$\mathcal{L}=21$ pb$^{-1}$. In the experimental analysis, di-jets events were 
selected by choosing the two jets with the highest transverse momentum 
$p_{T;1,2}$ from a single event that fell in the range 
$-0.8\leq\eta_{1,2}\leq 0.8$, with $\eta_{1,2}$ the rapidity of each of the two 
jets. Furthermore, an asymmetric condition was placed on the transverse 
momentum of each of the two jets, specifically $p_{T;1}\geq 8.0$ GeV and 
$p_{T;2}\geq 6.0$ GeV. The jet reconstruction procedures followed those used in 
the previous single-inclusive jet analysis~\cite{Adamczyk:2014ozi}, 
including the usage of the anti-$k_T$ algorithm~\cite{Cacciari:2008gp}
for jet finding (with resolution parameter $R=0.6$).
Asymmetries are provided as a function of the invariant mass of the di-jet 
pair, $M^{2\rm jets}$, for two distinct topologies in seven bins each: 
{\it same-sign}, in which both jets have either positive or negative 
rapidities, and {\it opposite-sign}, in which one jet has positive and the 
other has negative rapidity. The opposite-sign topology selects events arising 
from relatively symmetric (in $x$) partonic collisions, whereas same-sign
events select more asymmetric collisions. Using LO kinematics, one can expect 
this data to constrain the polarized gluon PDF with 
$0.01\lesssim x \lesssim 0.2$ and $\mu\sim M^{2\rm jets}$, where 
$15$ GeV $\lesssim M^{2\rm jets}\lesssim 70$ GeV. 
The uncertainties on the data are dominated by statistics. A full
breakdown of systematics is also provided, along with the correlation matrix
for the quadrature sum of the point-to-point statistical and systematic 
uncertainties between single- and di-jet asymmetries from the 2009 run. 
There are two additional systematics, which are $100\%$ correlated among bins,
originating from the beam polarization ($6.5\%$ of the asymmetry) and from
the relative luminosity (an offset of $5\times 10^{-4}$ on the asymmetry).

Third, I consider measurements of the double-spin asymmetry for mid-rapidity
neutral pion production, $A_{LL}^{\pi^0}$, recorded by the PHENIX experiment
during the 2009 run at $\sqrt{s}=200$ GeV (and combined with previous 
2005 and 2006 runs at the same center-of-mass energy)~\cite{Adare:2014hsq},
and the 2012-2013 runs at $\sqrt{s}=510$ GeV~\cite{Adare:2015ozj}. 
The integrated luminosity of the samples are $\mathcal{L}=6.5$ pb$^{-1}$
(for the 2009 run), $\mathcal{L}=20$ pb$^{-1}$ (for the 2012 run) and 
$\mathcal{L}=108$ pb$^{-1}$ (for the 2013 run). The data is presented in 
ten bins, for the asymmetries at $\sqrt{s}=200$ GeV, and in fourteen bins,
for the asymmetries at $\sqrt{s}=510$ GeV, as a function of the transverse 
momentum of the final neutral pion, $p_T^{\pi^0}$. The rapidity coverage of
the detector is $|\eta|<0.35$. Using LO kinematics, one can expect this data
to constrain the polarized gluon PDF with $0.05\lesssim x\lesssim 0.2$ 
when $\sqrt{s}=200$ GeV ($0.01\lesssim x\lesssim 0.2$ when $\sqrt{s}=510$ GeV)
and $\mu\sim p_T^{\pi^0}$, where $1$ GeV $\lesssim p_T^{\pi^0}\lesssim 17$ GeV.
The uncertainties on the data are dominated by statistics, except for
small-$p_T^{\pi^0}$ bins. There are two bin-by-bin fully correlated systematics
originating from the beam polarization ($4.8\%$ and $6.5\%$ of the asymmetry
at $\sqrt{s}=200$ GeV and $\sqrt{s}=510$ GeV respectively) and from the
relative luminosity (an offset of $4.2\times 10^{-4}$ and $3.6\times 10^{-4}$
on the asymmetry at $\sqrt{s}=200$ GeV and $\sqrt{s}=510$ GeV respectively).

Here, I do not consider other recent measurements performed at RHIC,
specifically neutral pion production asymmetries measured by STAR at 
$\sqrt{s}=200$ GeV~\cite{Adamczyk:2013yvv} and $\sqrt{s}=510$ 
GeV~\cite{Dilks:2016rkc}, and $W$-boson production asymmetries measured by 
PHENIX at $\sqrt{s}=500$ GeV and $\sqrt{s}=510$ GeV~\cite{Adare:2015gsd}.
This data is affected by rather large uncertainties, which are likely to 
limit its impact on PDFs~\cite{Nocera:2014gqa}. 
The data sets considered in this analysis 
are summarized in Tab.~\ref{tab:dataset}, where I indicate, for each of them, 
the name adopted to denote them, their corresponding publication 
reference, the integrated luminosity, $\mathcal{L}$, the center-of-mass 
energy, $\sqrt{s}$, the measured asymmetry, $\mathcal{A}$, and the number of 
data points, $N_{\rm dat}$.

\begin{table}[t]
\caption{The data sets considered in this analysis. For each set, I indicate
the name adopted to denote it, the corresponding publication reference,
the integrated luminosity, $\mathcal{L}$, the center-of-mass energy, $\sqrt{s}$,
the measured asymmetry, $\mathcal{A}$,  and the number of data points, 
$N_{\rm dat}$.}
\centering
\ruledtabular
\begin{tabular}{lccccc}
Data set & Ref. 
& $\mathcal{L}$ [pb$^{-1}$] & $\sqrt{s}$ [GeV] & $\mathcal{A}$ & $N_{\rm dat}$\\ 
\hline\\[-6pt]
STAR13-$W^-$ & \cite{Gunarathne:2017fnd} 
& 246.2 & 510 & $A_{L}^{e^-}$ & 4\\
STAR13-$W^+$ & \cite{Gunarathne:2017fnd} 
& 246.2 & 510 & $A_{L}^{e^+}$ & 4\\[2pt]
\hline\\[-6pt]
STAR09-2j-ss & \cite{Adamczyk:2016okk} 
& 21 & 200 & $A_{LL}^{2\rm jets}$ & 7\\[2pt]
STAR09-2j-os & \cite{Adamczyk:2016okk} 
& 21 & 200 & $A_{LL}^{2\rm jets}$ & 7\\[2pt]
\hline\\[-6pt]
PHENIX09-$\pi^0$ & \cite{Adare:2014hsq} 
& 6.5 & 200 & $A_{LL}^{\pi^0}$ & 12\\[2pt]
PHENIX13-$\pi^0$ & \cite{Adare:2015ozj} 
& 128 & 510 & $A_{LL}^{\pi^0}$ & 14\\
\end{tabular}
\ruledtabular
\label{tab:dataset}
\end{table}

\begin{figure}[t]
\centering
\includegraphics[scale=0.17,angle=270]{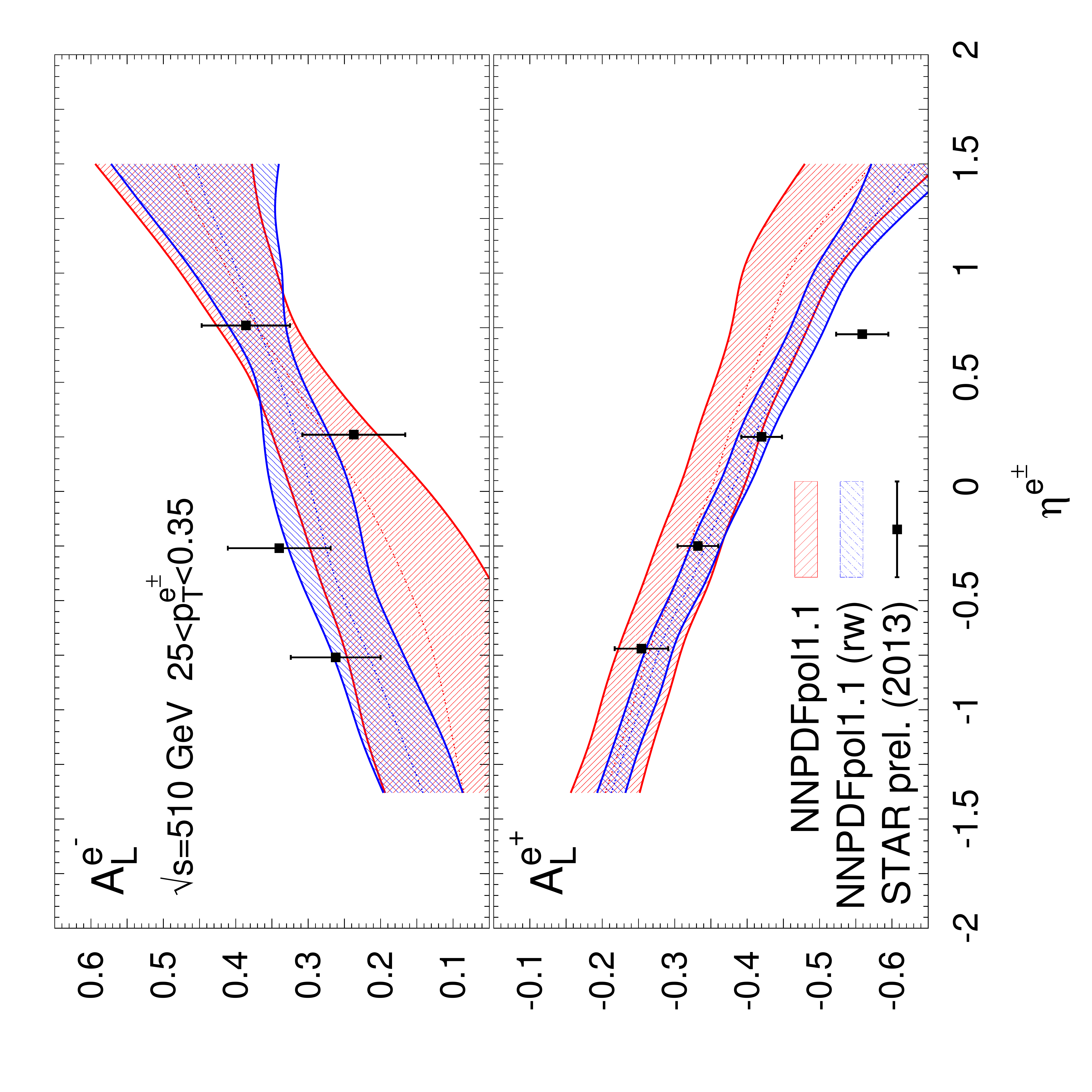}\\
\caption{The electron (top) and positron (bottom)
single-spin asymmetry for $W^-$ and $W^+$ production in single-polarized 
proton-proton collisions at $\sqrt{s}=510$ GeV, as a function of the decay 
electron/positron rapidity $\eta^{e^\pm}$. The data is from 
preliminary STAR measurements taken during the 2013 
run~\cite{Gunarathne:2017fnd}. Displayed data uncertainties are statistical 
only. Theoretical predictions are computed at NLO with the NNPDFpol1.1 
polarized PDF set and the NNPDF3.0 unpolarized PDF set. Results are shown 
before and after reweighting with the corresponding data. One-sigma uncertainty 
bands reflect uncertainties from the input polarized PDFs only.}
\label{fig:Ws}
\end{figure}

\begin{figure}[t]
\centering
\includegraphics[scale=0.17,angle=270]{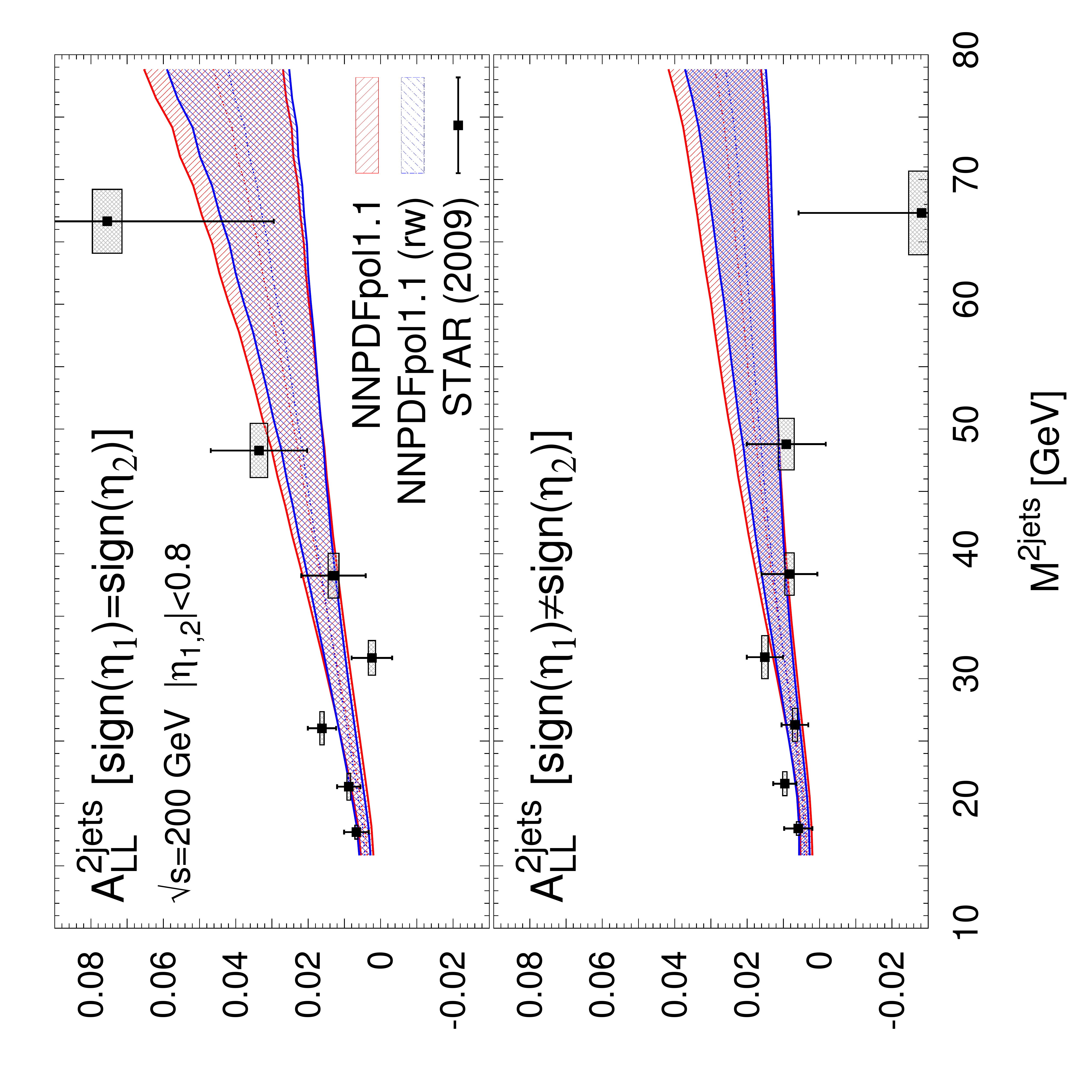}\\
\caption{The double-spin asymmetry with same-sign (top) and opposite-sign
(bottom) rapidity topologies for di-jet production in double-polarized
proton-proton collisions at $\sqrt{s}=200$ GeV, as a function of the invariant
mass of the di-jet pair, $M^{2{\rm jets}}$. The experimental data is from STAR
measurements taken during the 2009 run~\cite{Adamczyk:2016okk}. Statistical and 
systematic (shaded boxes) data uncertainties are displayed. Theoretical 
predictions and uncertainties bands are shown as in Fig.~\ref{fig:Ws}.}
\label{fig:dijets}
\end{figure}

\begin{figure}[t]
\centering
\includegraphics[scale=0.17,angle=270]{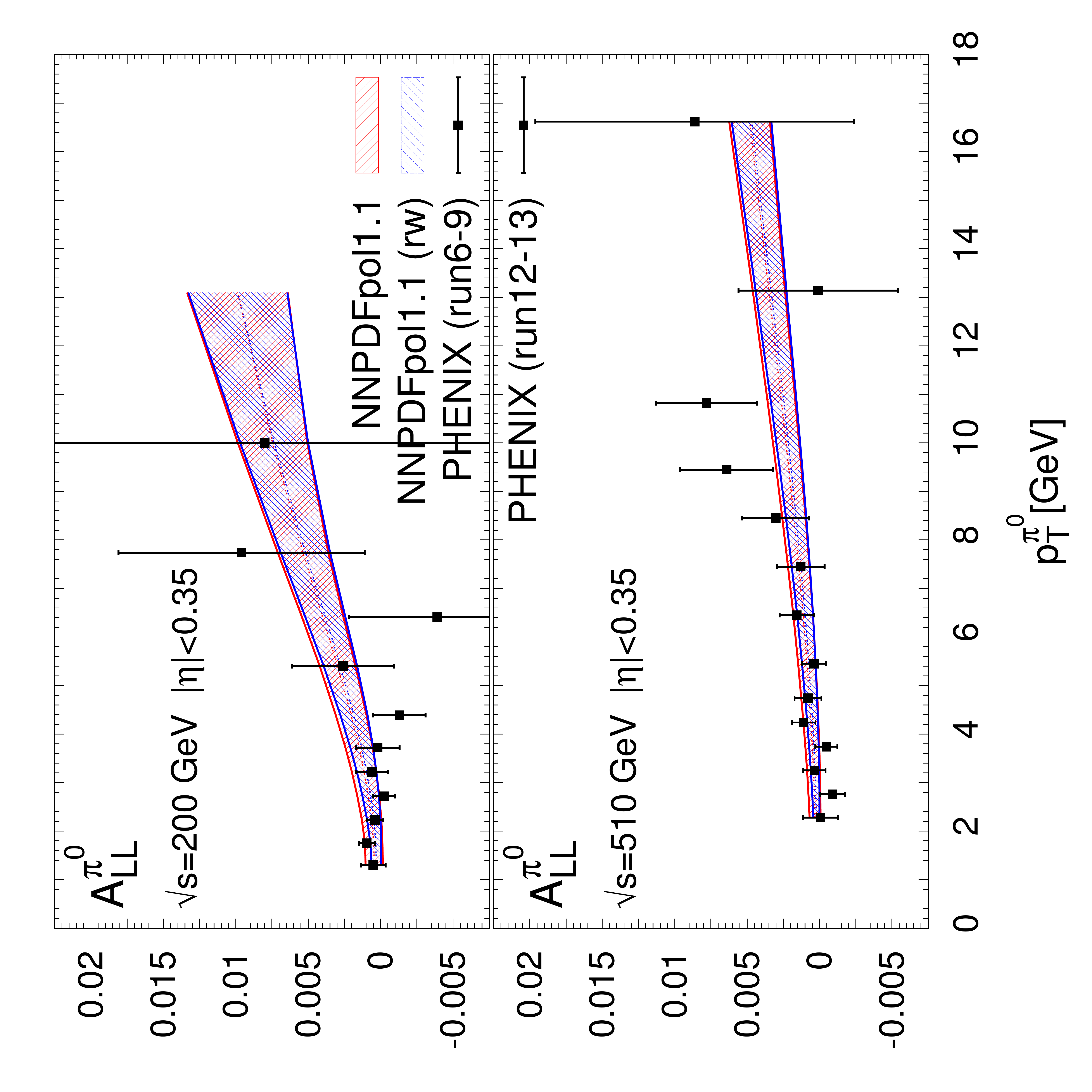}\\
\caption{The double-spin asymmetry for neutral pion production in 
double-polarized proton-proton collisions at $\sqrt{s}=200$ GeV from PHENIX
measurements taken during the 2005-2006-2009 runs (top) and at 
$\sqrt{s}=510$ GeV from PHENIX measurements taken during the 2012-2013 runs 
(bottom). Results are displayed as a function of the neutral pion transverse 
momentum $p_T^{\pi^0}$. Displayed data uncertainties are statistical only.
Theoretical predictions and uncertainties bands are shown as in 
Fig.~\ref{fig:Ws}. The NNFF1.0 set of FFs~\cite{Nocera:2017qgb} is used.}
\label{fig:pi0}
\end{figure}

In Figs.~\ref{fig:Ws}-\ref{fig:dijets}-\ref{fig:pi0}, I compare the 
measured asymmetries for each of the data sets collected in 
Tab.~\ref{tab:dataset} with the corresponding theoretical predictions.
The latter have all been computed at NLO accuracy in perturbative QCD.
They have been obtained using the NNPDFpol1.1 parton set~\cite{Nocera:2014gqa} 
for polarized PDFs and the NNPDF3.0 parton set~\cite{Ball:2014uwa} for 
unpolarized PDFs. In the case of neutral pion production asymmetries, 
the preliminary NNFF1.0 parton set~\cite{Nocera:2017qgb} has been used for FFs.
All used sets are NLO. The uncertainties on the theoretical 
predictions displayed in Figs.~\ref{fig:Ws}-\ref{fig:dijets}-\ref{fig:pi0} 
are one-sigma bands. They have been computed by propagating the uncertainty 
on the polarized PDFs only, while unpolarized PDFs and FFs have been set to 
their central values. This is justified because, in the kinematic region where 
asymmetries are measured, the unpolarized PDF and FF uncertainties are 
subdominant with respect to the polarized PDF uncertainty.
Furthermore, in the case of neutral pion production asymmetries, I have
explicitly checked that almost indistinguishable predictions are obtained
with either the NNFF1.0 or the DSS15~\cite{deFlorian:2014xna} FF sets.
Theoretical predictions have been computed by using the following pieces of 
code, which have been modified to handle NNPDF parton sets.
For $W$-boson production, I have used the code of Ref.~\cite{deFlorian:2010aa};
for di-jet production, I have used the code of Ref.~\cite{deFlorian:1998qp},
supplemented with the FastJet Package~\cite{Cacciari:2011ma};
for neutral pion production, I have used the code of Ref.~\cite{Jager:2002xm}.


The agreement between the data and theoretical predictions is quantified by the 
$\chi^2$ per data point, $\chi^2/N_{\rm dat}$, which is displayed in the 
third column of Tab.~\ref{tab:reweighting}. This is computed by taking into 
account all the available information on statistical and sytematic 
uncertainties, including their correlations, when the experimental covariance 
matrix is constructed. From Tab.~\ref{tab:reweighting} and 
Figs.~\ref{fig:Ws}-\ref{fig:dijets}-\ref{fig:pi0}, NLO theoretical predictions
based on NNPDF parton sets do appear to provide a good description of 
di-jet and neutral pion production double-spin asymmetries, while they do not
for $W$-boson production single-spin asymmetries. This is not unexpected.
On the one hand, the NNPDFpol1.1 set already included jet production data, 
of which di-jet production data are a subsample. They allowed for a rather
accurate determination of the gluon PDF in the same kinematic region where
it is sensitive to di-jet and neutral pion production data. On the other
hand, the NNPDFpol1.1 set included the STAR 2010-2011 $W$-boson production data,
which had significantly larger uncertainties than the new STAR 2013 data. 
For di-jet and neutral pion production asymmetries, the size of the 
uncertainties on theoretical predictions is comparable to (or smaller than) 
the size of the statistical uncertainties on the data. Conversely, for 
$W$-boson production asymmetries, theoretical uncertainties are always larger
than data uncertainties. One should then expect the largest impact on the 
underlying PDFs from the last set of data.  

\begin{table}[t]
\caption{The values of the $\chi^2$ per data point, $\chi^2/N_{\rm dat}$
($\chi^2_{\rm rw}/N_{\rm dat}$), before (after) reweighting, the effective 
number of replicas, $N_{\rm eff}$, and the modal value of the 
$\mathcal{P}(\alpha)$ distribution, $\langle\alpha\rangle$, for each data set 
with which the NNPDFpol1.1 PDFs are reweighted. The number of data points for 
each data set, $N_{\rm dat}$, is displayed for reference.}
\centering
\ruledtabular
\begin{tabular}{lccccc}
Data set & $N_{\rm dat}$ & $\chi^2/N_{\rm dat}$ & $\chi^2_{\rm rw}/N_{\rm dat}$ 
& $N_{\rm eff}$ & $\langle\alpha\rangle$\\
\hline\\[-6pt]
STAR13-$W^-$ & 4 & 2.44 & 0.69 & 38 & 1.35\\
STAR13-$W^+$ & 4 & 3.08 & 1.30 & 29 & 1.55\\[2pt]
\hline\\[-6pt]
STAR09-2j-ss & 7 & 1.41 & 1.18 & 89 & 1.10\\[2pt]
STAR09-2j-os & 7 & 1.26 & 0.83 & 92 & 1.05\\[2pt]
\hline\\[-6pt]
PHENIX09-$\pi^0$ & 12 & 0.69 & 0.60 & 84 & 0.75\\[2pt]
PHENIX13-$\pi^0$ & 14 & 0.61 & 0.58 & 92 & 0.80\\
\end{tabular}
\ruledtabular
\label{tab:reweighting}
\end{table}

In order to assess the impact of the new RHIC data on the 
polarized PDFs, I include them in the NNPDFpol1.1 parton set by means of 
Bayesian reweighting~\cite{Ball:2010gb,*Ball:2011gg}. This methodology 
consists in updating the representation of the probability distribution in the
space of PDFs - provided by a {\it prior} Monte Carlo ensemble of equally
probable PDFs - by means of Bayes' theorem. Specifically, each replica in the 
prior set is assigned a weight which assesses the probability that this replica 
agrees with the new data. The weights are computed by evaluating
the $\chi^2$ of the new data to the prediction obtained using a given replica.
After reweighting, replicas with small weights become almost irrelevant 
in ensemble averages, and the number of effective replicas in the Monte Carlo 
ensemble (see Eq.~(10) in Ref.~\cite{Ball:2010gb}) is smaller than the starting 
one. The consistency of the data used for reweighting with that 
included in the prior set can be assessed by examining the $\chi^2$ profile of 
the new data, $\mathcal{P}(\alpha)$, where $\alpha$ is the factor by which
the uncertainty on the new data must be rescaled in order for both 
the prior and the reweighted sets to 
be consistent with each other (see Eq.~(12) in Ref.~\cite{Ball:2010gb}). 
If the modal value of $\alpha$ is close to unity, the new data is consistent 
with the old, and its uncertainties have been correctly estimated.

I perform a simultaneous reweighting of the NNPDFpol1.1 parton set 
with all the data sets listed in Tab.~\ref{tab:dataset}. In 
Tab.~\ref{tab:reweighting}, I show the values of the $\chi^2$ per data point 
after reweighting, $\chi^2_{\rm rw}/N_{\rm dat}$, the number of effective replicas, 
$N_{\rm eff}$, and the modal value of the $\mathcal{P}(\alpha)$ distribution, 
$\langle\alpha\rangle$. The corresponding asymmetries, after reweighting 
with the new data, are displayed in 
Figs.~\ref{fig:Ws}-\ref{fig:dijets}-\ref{fig:pi0}, on top of their 
counterparts before reweighting.

The value of the $\chi^2$ per data point always decreases after reweighting. 
The improvement is marked for the $W$-boson production data, moderate for the 
di-jet production data and only slight for the neutral pion production data. 
This is expected, since the first data set has the smallest uncertainties, 
among all, in comparison to the PDF uncertainties on the theoretical 
predictions. This suggests that this data is bringing in a significant amount 
of new information. After reweighting, the $\chi^2$ per data point 
is of order one for all the new data sets. However, in the case of 
$W$-boson and neutral pion production asymmetries, these numbers should be 
taken with care, because a complete information on correlated systematics is 
not available. This is the reason why the reweighted $\chi^2$ is smaller than 
one for these sets, except for the $W^+$ production data. In this case, 
the value of the $\chi^2$ is raised by a sizable contribution coming from the 
point with the largest positron rapidity, which disagrees by about two sigma 
with the reweighted theoretical prediction and the previous STAR measurement
from run 2010-2011 (see also Fig.~4 in ref~\cite{Gunarathne:2017fnd}). 

The number of effective replicas after reweighting depends significantly on 
the data set. The size of the reweighted parton set is about $90\%$ of
the original NNPDFpol1.1 parton set (made of $N_{\rm rep}=100$ replicas) 
for di-jet and pion production data, while it is only about $30\%$-$40\%$ 
for $W$-boson production data. This result reflects the different constraining 
power of the various data sets, which is maximized in the last case. 
In principle, a prior ensemble with a larger number of replicas should then 
be needed for the reweighted ensemble to sample the probability density in the
space of PDFs with as much accuracy. However this is not relevant here, as
reweighted results only serve to assess the impact of the new data, and are not
used to construct a new parton set.

The modal value of the $\mathcal{P}(\alpha)$ distribution is of order one
for all the new data sets. Values of $\langle\alpha\rangle$ slightly 
larger than one are found for the $W$-boson production data: in the case
of $W^-$, this is mostly determined by a sizable fluctuation of one data 
point (around $\eta^{e^-}\sim 0.25$) with respect to the shape of the 
corresponding asymmetry; in the case of $W^+$, this is mostly determined by the 
fourth data point (around $\eta^{e^+}\sim 0.75$), which, as already noted, 
disagrees by about two sigma with the reweighted theoretical prediction.
Values of $\langle\alpha\rangle$ slightly smaller than one are found for the
neutral pion production data, thus suggesting that experimental uncertainties 
are likely to be overestimated, possibly because of the lack of a 
complete information on correlations among systematics.

\begin{figure}[t]
\centering
\includegraphics[width=\columnwidth,angle=270,clip=true,trim=6cm 6cm 6cm 6cm]{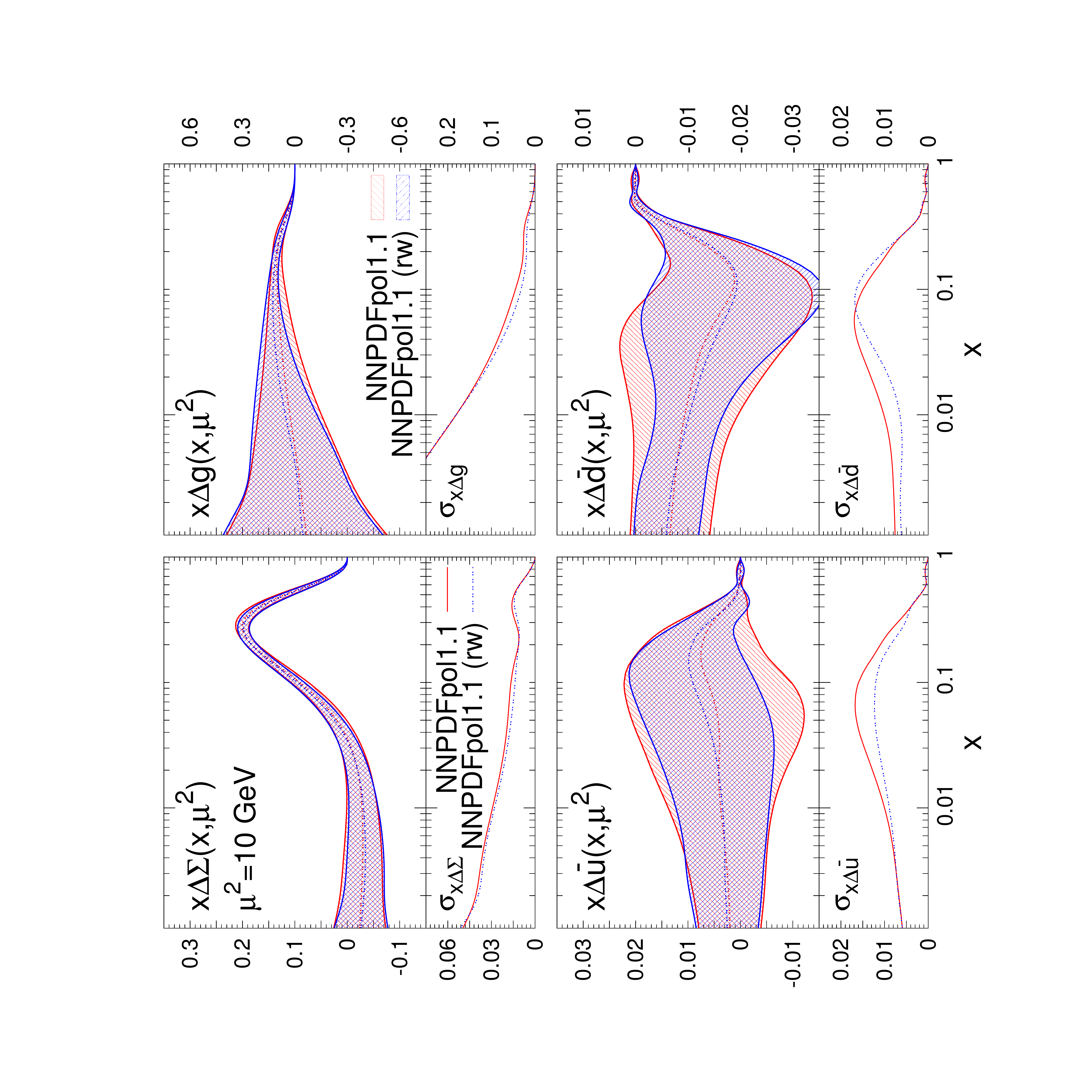}
\caption{A comparison of polarized PDFs before and after the
simultaneous reweighting of NNPDFpol1.1 with the data sets listed in 
Tab.~\ref{tab:dataset}. From left to right, top to bottom, the 
singlet, $\Delta\Sigma$, the gluon, $\Delta g$, and the up and down sea quarks, 
$\Delta\bar{u}$ and $\Delta\bar{d}$, are displayed. Parton distributions are 
evaluated at $\mu^2=10$ GeV. Bands represent one-sigma uncertainties; 
they are also shown in the lower inset of each panel.}
\label{fig:pdfs}
\end{figure}

In Fig.~\ref{fig:pdfs}, I compare the polarized PDFs before and after 
the simultaneous reweighting of NNPDFpol1.1 with the data sets listed in 
Tab.~\ref{tab:dataset}. From left to right, top to bottom, I show the 
singlet (for $n_f$ active flavors), 
$\Delta\Sigma=\sum_{i=1}^{n_f}(\Delta q_i + \Delta\bar{q}_i)$,
the gluon, $\Delta g$, and the up and down sea quarks, $\Delta\bar{u}$ and 
$\Delta\bar{d}$. Parton distributions are evaluated at $\mu^2=10$ GeV$^2$.
Bands represent one-sigma uncertainties, which are also displayed 
separately in the lower inset of each panel. 

The impact of the wew data on the polarized PDFs of the proton is twofold.
On the one hand, it induces a shift of the PDF central values, as a 
consequence of the adjustment to the shape of the corresponding asymmetries. 
Specifically, the central value of $\Delta g$ increases by about $30\%$ of its
original value in the region $0.1\lesssim x\lesssim 0.2$; the central value of 
$\Delta\bar{u}$ increases by about $25\%$ and that of $\Delta\bar{d}$ decreases 
by about $10\%$ approximately in the same region of $x$ wher also 
$\Delta g$ is affected. On the other hand, 
the new data induces a reduction of the PDF uncertainties, as a 
consequence of the improved precision of the corresponding asymmetries.
For $\Delta g$ and $\Delta\bar{d}$ such a reduction is moderate, and not larger 
than $5\%$ of its original value; for $\Delta\bar{u}$ it is fairly more 
pronounced, around $25\%$. As expected, these differences appear in the
kinematic region where the PDFs are sensitive to the new data,
roughly $0.1\lesssim x\lesssim 0.2$.
Central values and uncertainties of other PDFs,
like the singlet or the up, down and strange (not shown in Fig.~\ref{fig:pdfs}) 
are not significantly affected by the new data.

The impact of the new data is also revealed by the truncated first moments of 
the corresponding PDFs in the region $[x_{\rm min}, x_{\rm max}]$
\begin{equation}
\langle\Delta f(\mu^2) \rangle^{[x_{\rm min}, x_{\rm max}]}
\equiv
\int_{x_{\rm min}}^{x_{\rm max}}dx\,\Delta f(x,\mu^2)
\,\mbox{.}
\label{eq:trmoments}
\end{equation}
Specifically, Eq.~(\ref{eq:trmoments}) gives, at $\mu^2=Q^2=10$ GeV$^2$,
respectively before and after reweighting: 
$\langle\Delta g(Q^2)\rangle^{[0.01,0.2]}=0.23\pm 0.23$ and 
$\langle\Delta g(Q^2)\rangle^{[0.01,0.2]}_{\rm rw}=0.32\pm 0.21$;
$\langle\Delta \bar{u}(Q^2)\rangle^{[0.01,0.2]}=0.01\pm 0.04$ and 
$\langle\Delta \bar{u}(Q^2)\rangle^{[0.01,0.2]}_{\rm rw}=0.02\pm 0.03$;
$\langle\Delta \bar{d}(Q^2)\rangle^{[0.01,0.2]}=-0.04\pm 0.04$ and 
$\langle\Delta \bar{d}(Q^2)\rangle^{[0.01,0.2]}_{\rm rw}=-0.05\pm 0.03$.

In summary, in comparsion to the NNDPFpol1.1 analysis, a slightly more positive 
polarized gluon PDF, extended to somewhat smaller values of momentum 
fractions with narrower uncertainties, as well as a slightly more marked and
more precise asymmetry between polarized up and down sea quarks 
are found in this study. These results confirm the importance of 
the RHIC spin physics program in the understanding of the longitudinal spin 
structure of the proton.

\begin{acknowledgments}

I would like to thank the organizers of the SPIN2016 conference for the
opportunity to be a convener of Working Group B: {\it Nucleon Helicity 
Structure} and for financial support. I am also grateful to Sasha Bazilevsky, 
for discussions on neutral pion production data from the PHENIX experiment,
and to Brian Page, for clarifications on systematic 
uncertainties of di-jet production data from the STAR experiment.
This work is supported by a STFC Rutherford Grant ST/M003787/1.
 
\end{acknowledgments}

\nocite{*}
\bibliography{NOCERA_proceedings_2}

\end{document}